\documentclass[12pt]{article}

\usepackage[normalem]{ulem}
\usepackage[mathscr]{eucal}
\usepackage[version=3]{mhchem}
\usepackage{stmaryrd}

\usepackage{epsfig,latexsym,amsfonts,amsmath,amsthm,amssymb,amsbsy,multirow,slashed,wasysym,textcomp,dsfont,comment,mathtools,cancel,cite,datetime,appendix}
\usepackage{tocloft}
\usepackage{bbold}
\usepackage{tikz}
\usetikzlibrary{decorations.pathreplacing,calligraphy}
\usetikzlibrary{backgrounds}
\usetikzlibrary{patterns}
\usetikzlibrary{arrows.meta}
\tikzstyle{bag} = [align=center]
\usetikzlibrary{decorations.pathmorphing}
\usepackage{hyperref}
\usepackage{subcaption}

\usepackage{array}
\usepackage{booktabs}

\newcommand{\badat}{\begin{alignedat}}
\newcommand{\eadat}{\end{alignedat}}
\newcommand\scalemath[2]{\scalebox{#1}{\mbox{\ensuremath{\displaystyle #2}}}}
\def\be{\begin{equation}}
\def\ee{\end{equation}}
\def\ba{\begin{aligned}}
\def\ea{\end{aligned}}

\usepackage{xcolor}

\newcommand{\pink}[1]{\textcolor{\pink}{#1}}

\definecolor{dblue}{rgb}{0.2,0.50,0.80}

\def\bz{{\bar z}}
\def\bw{{\bar w}}

\def\oh{\mathcal O}

\def\pa{{\partial}}
\def\a{{\alpha}}
\def\D{{\Delta}}
\def\d{{\delta}}

\def\n{{\nu}}
\def\s{{\sigma}}

\def\oh{{\mathcal O}}

\makeatletter
\newsavebox{\@brx}
\newcommand{\llangle}[1][]{\savebox{\@brx}{\(\m@th{#1\langle}\)}%
  \mathopen{\copy\@brx\kern-0.5\wd\@brx\usebox{\@brx}}}
\newcommand{\rrangle}[1][]{\savebox{\@brx}{\(\m@th{#1\rangle}\)}%
  \mathclose{\copy\@brx\kern-0.5\wd\@brx\usebox{\@brx}}}
\makeatother

\DeclareFontFamily{OT1}{pzc}{}
\DeclareFontShape{OT1}{pzc}{m}{it}{<-> s * [1.10] pzcmi7t}{}
\DeclareMathAlphabet{\mathpzc}{OT1}{pzc}{m}{it}

\definecolor{vert}{rgb}{0.1367 0.543 0.1367}

\usepackage[retainorgcmds]{IEEEtrantools}
\def\bea{\begin{IEEEeqnarray*}}
\def\eea{\end{IEEEeqnarray*}}
\def\n{\IEEEyesnumber}
\def\sn{\IEEEyessubnumber}

\numberwithin{equation}{section} 

\begin{document}

 \begin{titlepage}
  \thispagestyle{empty}
  \begin{flushright}
  \end{flushright}
  \bigskip
  \begin{center}

        \baselineskip=13pt {\LARGE {Multiparticle Contributions to the Celestial OPE
       }}

      \vskip1cm 

   \centerline{ 
   {Alfredo Guevara},${}^{1,{}2}$ 
    {Yangrui Hu},${}^3$ and {Sabrina Pasterski}\,${}^3$
}

\bigskip\bigskip

\centerline{\em${}^1$ 
\it Center for the Fundamental Laws of Nature, Harvard University, Cambridge, MA 02138
}

\bigskip

\centerline{\em${}^2$ 
\it Society of Fellows, Harvard University, Cambridge, MA 02138
}

\bigskip

 \centerline{\em${}^3$ 
\it Perimeter Institute for Theoretical Physics, Waterloo, ON N2L 2Y5, Canada}

\bigskip\bigskip

\end{center}

\begin{abstract}
 \noindent 

We start by defining two-particle operators that appear in celestial CFT. We then show how to compute their OPE coefficients with the known single-particle operators at tree level from multiparticle factorization channels, focusing on the leading contribution involving the two-particle states. These factorization channels only give us single-particle exchanges. To extract the multiparticle exchanges, we look at the $\overline{\rm MHV}$ gluon amplitudes and show how non-factorization channels contribute to two-particle terms in the single-helicity sector. This is a first step towards systematically computing the full celestial OPE.
  
\end{abstract}

\end{titlepage}

\tableofcontents

\section{Introduction}\label{sec:intro}

Celestial holography posits that scattering amplitudes are naturally encoded in a codimension-two CFT living on the celestial sphere. While this paradigm is motivated by the symmetry structure of asymptotically flat spacetimes, it suggests an interesting perspective of extracting the full ${\cal S}$-matrix from the collinear limits of scattering. 
If we want to follow this logic and set up such a celestial bootstrap program, the input we need is the full celestial CFT data -- the OPE coefficients and spectrum. 

The single-particle celestial operators were systematically studied in~\cite{Pasterski:2017kqt,Narayanan:2020amh,Muck:2020wtx,Law:2020tsg} and their OPE coefficients amongst each other were computed from amplitude splitting functions in~\cite{Pate:2019lpp,Fotopoulos:2019vac,Himwich:2021dau}. Now if we want our 2D theory to have a state-operator correspondence and for the 2D theory to be dual to our 4D asymptotically flat Hilbert space, we expect composite operators to capture the multiparticle states. These will be classified in~\cite{Kulp:2024scx}, but we have already seen how these affect interpretations of the associativity of the celestial symmetry generators in~\cite{Ball:2023sdz}. Here we perform the first steps towards systematically computing the full celestial OPE.

Let us start by defining the following composite two-particle celestial operators
\begin{equation}
    \begin{split}
        :\oh_1^{(n_1,m_1)}\oh^{(n_2,m_2)}_2: (w,\bw) ~\equiv&~ \oint \frac{dz}{2\pi i}\frac{1}{z-w}\,\oint \frac{d\bz}{2\pi i}\frac{1}{\bz-\bw}\\
        &\qquad \pa_z^{n_1}\pa_{\bz}^{m_1}\oh_{\D_1,J_1}(z,\bz)\,\pa_w^{n_2}\pa_{\bw}^{m_2}\oh_{\D_2,J_2}(w,\bw) ~~.
    \end{split}
\end{equation}
By taking appropriate linear combinations, one can extract a basis of 2D primaries and their descendants~\cite{Kulp:2024scx}\,\footnote{Note that when considering non-gravitational bulk theory, ``primaries" and ``descendants" are defined with respect to the global conformal group, as adopted in this paper.}. By matching the scaling dimension and spin, we see that in a celestial OPE, these terms contribute as follows
\begin{equation}
    \oh_{\D_1,J_1}(z,\bz)\,\oh_{\D_2,J_2}(w,\bw) ~\sim~ (z-w)^{n_1+n_2}\,(\bz-\bw)^{m_1+m_2}\,:\oh_1^{(n_1,m_1)}\oh^{(n_2,m_2)}_2: (w,\bw)~~.
\end{equation}
Note that $n_1$, $n_2$, $m_1$, and $m_2$ are all non-negative integers, so the most leading term in the OPE is $n_1=n_2=m_1=m_2=0$, and will be our focus in this paper. This term is indeed a celestial primary with weight $\Delta_1+\Delta_2$ and spin $J_1+J_2$. To simplify the notation in what follows, we omit the $(0,0)$ superscripts and define  
\begin{equation}
     :\oh_1\oh_2:(w,\bw) ~\equiv~  \oint \frac{dz}{2\pi i}\frac{1}{z-w}\,\oint \frac{d\bz}{2\pi i}\frac{1}{\bz-\bw}\,\oh_1(z,\bz)\,\oh_2(w,\bw)~~.
    \label{eq:def}
\end{equation}  
We will focus on gluon amplitudes in this paper and our goal will be to extract the OPE coefficients between the following operators
\begin{equation}
    \begin{split}
    &\text{single-particle operators:}~~\oh^{a,-}(z,\bz) ~~,~~
    \oh^{a,+}(z,\bz) ~~,~~ \\
    &\text{two-particle operators:}~~:\oh^{a,-}\oh^{b,-}:(z,\bz) ~~,~~
    :\oh^{a,-}\oh^{b,+}:(z,\bz) ~~,~~
    :\oh^{a,+}\oh^{b,+}:(z,\bz) ~~,~~
    \end{split}
\end{equation}
where $:\oh^{a,+}\oh^{b,-}:(z,\bz)~=~:\oh^{b,-}\oh^{a,+}:(z,\bz)$. Note that the superscripts here are the helicities in all-out notation and, for simplicity, all of our particles are outgoing.

From the scaling dimension and spin, we expect the following ansatz for the OPEs
\bea{l}\label{equ:OOtoO} \n
    \oh_1^{a,-}(z_1,\bz_1)\,\oh_2^{b,-}(z_2,\bz_2) ~\sim~ \frac{{\cal F}^{a^-b^-}{}_{c^-}}{\bz_{12}}\,\oh^{c,-}_q(z_2,\bz_2) ~+~ :\oh^{a,-}_1\oh_2^{b,-}:(z_2,\bz_2) ~+~\cdots~~ ~~\sn \label{equ:O-O-toO-} \\
    \oh_1^{a,-}(z_1,\bz_1)\,\oh_2^{b,+}(z_2,\bz_2) ~\sim~ \frac{{\cal F}^{a^-b^+}{}_{c^-}}{z_{12}}\,\oh^{c,-}_q(z_2,\bz_2) ~+~ \frac{{\cal F}^{a^-b^+}{}_{c^+}}{\bz_{12}}\,\oh^{c,+}_q(z_2,\bz_2) \sn \nonumber \\ 
    \qquad\qquad\qquad\qquad\qquad\qquad\qquad\qquad\qquad\qquad~~~ ~+~ :\oh^{a,-}_1\oh_2^{b,+}:(z_2,\bz_2) ~+~\cdots~~~~\sn \label{equ:O-O+toO-}\\
    \oh_1^{a,+}(z_1,\bz_1)\,\oh_2^{b,+}(z_2,\bz_2) ~\sim~ \frac{{\cal F}^{a^+b^+}{}_{c^+}}{z_{12}}\,\oh^{c,+}_q(z_2,\bz_2) ~+~ :\oh^{a,+}_1\oh_2^{b,+}:(z_2,\bz_2)~+~\cdots~~ ~~\sn \label{equ:O+O+toO+}
\eea
and
\bea{l}\label{eq:O:OO:OPE} \n
    \oh^{a,-}_1(z_1,\bz_1)\,:\oh^{b,-}_2\oh^{c,-}_3:(z_3,\bz_3)~\sim~\frac{{\cal C}^{a^-b^-c^-}{}_{d^-}}{\bz_{13}^2}\,\oh^{d,-}_p(z_3,\bz_3) \sn \nonumber \\
    \qquad\qquad\qquad\qquad\qquad\qquad\qquad\qquad~~~~ ~+~ \frac{{\cal D}^{a^-b^-c^-}{}_{d^-e^-}}{\bz_{13}}\,:\oh^{d,-}_p\oh^{e,-}_q:(z_3,\bz_3) ~+~ \cdots \sn\label{eq:O:OO:OPE---}\\
    \oh^{a,-}_1(z_1,\bz_1)\,:\oh^{b,-}_2\oh^{c,+}_3:(z_3,\bz_3) ~\sim~ \frac{{\cal C}^{a^-b^-c^+}{}_{d^+}}{\bz_{13}^2}\,\oh^{d,+}_p(z_3,\bz_3) + \text{two-particle terms} + \cdots     \sn\label{eq:O:OO:OPE--+}\\
    \oh^{a,-}_1(z_1,\bz_1)\,:\oh^{b,+}_2\oh^{c,+}_3:(z_3,\bz_3) ~\sim~ \frac{{\cal C}^{a^-b^+c^+}{}_{d^-}}{z_{13}^2}\,\oh^{d,-}_p(z_3,\bz_3) + \text{two-particle terms} + \cdots     \sn\label{eq:O:OO:OPE-++}
\eea
where other helicity configurations can be easily obtained by conjugation.\,\footnote{In the current work, we only consider the contributions written in (\ref{eq:O:OO:OPE}). The omitted terms will be subleading, as discussed in section~\ref{sec:universal}.}
The ${\cal F}$'s were computed in~\cite{Pate:2019lpp,Fotopoulos:2019vac,Himwich:2021dau} and we will compute ${\cal C}$'s and ${\cal D}$'s here. 
The coefficient ${\cal C}$ can be obtained for any gluon amplitude just from the 4-particle factorization channels.\,\footnote{Note that the single-particle exchanges in consecutive multi-gluon celestial OPEs have been discussed in~\cite{Ebert:2020nqf}.} 
While multiparticle exchanges will not appear in the factorization channels, we can extract their contributions from evaluating the full amplitudes. One thing worth noting is that the operators on the left- and right-hand sides of the OPEs are on a different footing until we understand the multiparticle inner products. This will be partially addressed in appendix~\ref{appen:2out}.

This paper is organized as follows. 
In section~\ref{sec:universal}, we compute the OPE coefficients ${\cal C}$  by taking multicollinear limits within the relevant 4-particle factorization channels in momentum space. By performing a Mellin transform to the boost basis, we obtain the Celestial OPE coefficients. These results hold for all tree-level gluon amplitudes. These factorization channels only capture single-particle exchanges with the rest of the celestial correlation function. Meanwhile, multiparticle exchanges receive contributions from non-factorization channels. In section~\ref{sec:2ptexchange} we show how to extract two-particle terms and constrain the OPE coefficient by evaluating a generic $n$-point $\overline{\rm MHV}$ gluon amplitude. Here we focus on the single-helicity sector as a guiding example. We close with several remarks in section~\ref{sec:conclusion}. Multicollinear limits and kinematics are reviewed in appendix~\ref{appen:conventions}, while we comment on the multiparticle generalization of the radially quantized $out$ states in appendix~\ref{appen:2out}.

\section{Single-particle Exchanges}\label{sec:universal}

In this section, we compute the single-particle exchanges appearing in the OPE between a single-particle operator and a two-particle operator. As mentioned above, the single-particle contributions can be derived solely from the factorization channels. Therefore, this result holds for generic $n$-point functions. In this work, we focus on tree-level gluon amplitudes and our procedure is as follows. 

First, note that the OPEs (\ref{eq:O:OO:OPE}) correspond to a specific consecutive multicollinear limit in the amplitude: particles 2 and 3 going collinear first and then particle 1. We can then consider (anti-)holomorphic multicollinear limit for $\oh_1$, $\oh_2$, and $\oh_3$, following~\cite{Ball:2023sdz}. 
Under such multicollinear limits, any $1/s_{12}$, $1/s_{13}$, $1/s_{23}$, and $1/s_{123}$ propagators go on-shell. We denote by $s_{ij}\sim \epsilon \to 0$ the rate at which we approach the limit. According to our ansatz \eqref{eq:O:OO:OPE---}, the double and single poles in the multiparticle OPE correspond to divergences $1/\epsilon^2$ and $1/\epsilon$ respectively. 
Now, for a general $n$-point amplitude, the leading (i.e. $1/\epsilon^2$) divergence is controlled by a splitting function entering via the following factorization 
\begin{equation}
    A_n[1^{h_1,a}2^{h_2,b}3^{h_3,c}\cdots n^{h_n,a_n}] ~\rightarrow~ \sum_{h_P=\pm}\,{\rm Split}\left[1^{h_1,a}2^{h_2,b}3^{h_3,c}\to P^{h_P,d}\right]\,A_{n-2}[P^{h_P,d}\cdots n^{h_n,a_n}] ~~.
    \label{equ:factorization}
\end{equation}
It is at this order that we can obtain single-particle exchange terms in the OPE, from general helicity amplitudes. Note that for the celestial correlators we should consider the full amplitude and not just the color-ordered ones. In the following we will employ the CSW rules \cite{Cachazo:2004kj} to derive the splitting function from the following $s$, $t$, and $u$ contributions:

\begin{equation}\resizebox{0.9\textwidth}{!}{$%
    \begin{aligned}
        & f^{abe}f^{ecd}\,\begin{tikzpicture}[baseline={([yshift=-0.9ex]current bounding box.center)},scale=0.9]
        \draw[thick] (-1,0) node[above right]{$-q$} --  (1,0) node[above left]{$q$};
        \draw[thick] (-1,0) -- (-1.5,1.73/2) node[left]{$1$};
        \draw[thick] (1,0) -- (1.5,1.73/2) node[right]{$-P$};
        \draw[thick] (-1,0) -- (-1.5,-1.73/2) node[left]{$2$};
        \draw[thick] (1,0) -- (1.5,-1.73/2) node[right]{$3$};
\end{tikzpicture} + f^{ace}f^{ebd}\,\begin{tikzpicture}[baseline={([yshift=-0.9ex]current bounding box.center)},scale=0.9]
        \draw[thick] (-1,0) node[above right]{$-q$} --  (1,0) node[above left]{$q$};
        \draw[thick] (-1,0) -- (-1.5,1.73/2) node[left]{$1$};
        \draw[thick] (1,0) -- (1.5,1.73/2) node[right]{$-P$};
        \draw[thick] (-1,0) -- (-1.5,-1.73/2) node[left]{$3$};
        \draw[thick] (1,0) -- (1.5,-1.73/2) node[right]{$2$};
\end{tikzpicture} +
f^{bce}f^{aed}\,\begin{tikzpicture}[baseline={([yshift=-0.9ex]current bounding box.center)},scale=0.9]
        \draw[thick] (0,-1+0.5) --  (0,1-0.5);
        \draw[thick] (0,-1+0.5) -- (1.73/2,-1.5+0.5) node[right]{$3$};
        \draw[thick] (0,0.5) -- (1.73/2,1.5-.5) node[right]{$-P$};
        \draw[thick] (0,-1+0.5) -- (-1.73/2,-1.5+0.5) node[left]{$2$};
        \draw[thick] (0,1-.5) -- (-1.73/2,1.5-.5) node[left]{$1$}; 
        \draw (0,0.4) node[right]{$q$};
        \draw (0,-0.4) node[right]{$-q$};
\end{tikzpicture}~~.
    \end{aligned}$}
    \label{equ:diagrams}
\end{equation}
According to the CSW prescription, each of these diagrams can be written as products of 3-point amplitudes and off-shell propagators. In this paper, we consider the case where the 3-point vertices in (\ref{equ:diagrams}) are either both MHV (choosing anti-holomorphic collinear limit) or both $\overline{\rm MHV}$ (choosing holomorphic collinear limit),
since it yields the most leading contributions of interest here.
This requirement leads to the following helicity configurations
\begin{equation}
{\rm Split}\left[1^{-,a}2^{-,b}3^{-,c}\to P^{-,d}\right]~~,~~
{\rm Split}\left[1^{-,a}2^{+,b}3^{+,c}\to P^{-,d}\right]~~,~~ {\rm Split}\left[1^{-,a}2^{-,b}3^{+,c}\to P^{+,d}\right]~~,~~
    \label{equ:helicity-config}
\end{equation}
and their conjugates and, in turn, to the ansatz in (\ref{eq:O:OO:OPE}). 
As mentioned earlier, in the RHS of (\ref{eq:O:OO:OPE}), there could be operators with different helicity, but these omitted terms are more subleading in the multicollinear expansion parametrized by $\epsilon$ (see appendix~\ref{appen:conventions} and \cite{Ball:2023sdz}).  As pointed out in~\cite{Birthwright:2005ak}, the splitting functions we are considering in (\ref{equ:helicity-config}) can be completely extracted from MHV or $\overline{\rm MHV}$ vertices, while for other configurations one needs to consider NMHV amplitudes.

Note that the coefficients in (\ref{equ:OOtoO}) and (\ref{eq:O:OO:OPE}) can be written in either the momentum basis as splitting functions, or the boost basis as celestial OPE coefficients. 
In the momentum basis, the operator $\oh_1$ carries energy $\omega_1$ while in the boost basis, it carries scaling dimension $\D_1$ and the two bases are related by a Mellin transform. 
The ${\cal F}$'s in momentum space are as follows 
\bea{l}\label{equ:OOtoOmom} \n
    \oh_1^{a,-}(z_1,\bz_1)\,\oh_2^{b,-}(z_2,\bz_2) ~\sim~ \frac{f^{abc}}{\bz_{12}}\,\frac{\omega_q}{\omega_1\omega_2}\,\oh^{c,-}_q(z_2,\bz_2) ~+~ \cdots ~~\sn \label{equ:O-O-toO-mom} \\
    \oh_1^{a,-}(z_1,\bz_1)\,\oh_2^{b,+}(z_2,\bz_2) ~\sim~ \frac{f^{abc}}{\bz_{12}}\,\frac{\omega_2}{\omega_1\omega_q}\,\oh^{c,+}_q(z_2,\bz_2) ~+~ \frac{f^{abc}}{z_{12}}\,\frac{\omega_1}{\omega_2\omega_q}\,\oh^{c,-}_q(z_2,\bz_2) ~+~ \cdots ~~\sn \label{equ:O-O+toO+mom}\\
    \oh_1^{a,+}(z_1,\bz_1)\,\oh_2^{b,+}(z_2,\bz_2) ~\sim~ \frac{f^{abc}}{z_{12}}\,\frac{\omega_q}{\omega_1\omega_2}\,\oh^{c,+}_q(z_2,\bz_2) ~+~ \cdots~~ ~~\sn \label{equ:O+O+toO+mom}
\eea
where $\omega_q=\omega_1+\omega_2$. 
We will compute the coefficients ${\cal C}$'s in the momentum basis in sections~\ref{sec:singleh-split} and \ref{sec:mixedh-split} before implementing a Mellin transform to the celestial basis in section~\ref{sec:Mellin-transf}.

\subsection{Single-helicity Splitting Functions}\label{sec:singleh-split}

In this section, we will focus on the single-helicity sectors. Namely, the helicity configurations $h_1=h_2=h_3=h_P=\pm$. 
Mechanically, the computations for both cases are similar, so we focus on the negative helicity one. 
In the CSW prescription a full amplitude can be constructed exclusively in terms of MHV amplitudes (i.e. two negative helicity states) continued to off-shell MHV vertices. For the case $h_i = -1$ it is clear that the three collinear particles must lie in different $(--+)$ vertices, and that the leading divergence is achieved when two of them lie in the same vertex, contiguous to the third particle. This results in the three diagrams (\ref{equ:diagrams}), which we evaluate as products of MHV vertices with an off-shell propagator, using the kinematics described in appendix~\ref{appen:conventions}. Then, the singular piece can be easily computed by plugging in the \textit{antiholomorphic} multicollinear limit, again following the conventions of the appendix. The result is 
\begin{align}
    \begin{tikzpicture}[baseline={([yshift=-0.9ex]current bounding box.center)},scale=0.9]
        \draw[thick] (-1,0) node[above right]{$-q^{+}$} --  (1,0) node[above left]{$q^-$};
        \draw[thick] (-1,0) -- (-1.5,1.73/2) node[left]{$1^-$};
        \draw[thick] (1,0) -- (1.5,1.73/2) node[right]{$-P^+$};
        \draw[thick] (-1,0) -- (-1.5,-1.73/2) node[left]{$2^-$};
        \draw[thick] (1,0) -- (1.5,-1.73/2) node[right]{$3^-$};
\end{tikzpicture} ~=&~ \frac{\langle12\rangle^3}{\langle2q\rangle\langle q1\rangle}\,\frac{1}{\langle12\rangle[12]}\,\frac{\langle q3\rangle^3}{\langle3P\rangle\langle Pq\rangle} ~=~\frac{\omega_P(\omega_1z_{13}+\omega_2z_{23})}{\omega_1\omega_2\bz_{12}} ~~,
\label{eq:s-diagram} \\
  \begin{tikzpicture}[baseline={([yshift=-0.9ex]current bounding box.center)},scale=0.9]
        \draw[thick] (0,-1+0.5) --  (0,1-0.5);
        \draw[thick] (0,-1+0.5) -- (1.73/2,-1.5+0.5) node[right]{$3^-$};
        \draw[thick] (0,0.5) -- (1.73/2,1.5-.5) node[right]{$-P^+$};
        \draw[thick] (0,-1+0.5) -- (-1.73/2,-1.5+0.5) node[left]{$2^-$};
        \draw[thick] (0,1-.5) -- (-1.73/2,1.5-.5) node[left]{$1^-$}; 
        \draw (0,0.4) node[right]{$q^-$};
        \draw (0,-0.4) node[right]{$-q^{+}$};
\end{tikzpicture} ~=&~ \frac{\langle23\rangle^3}{\langle3q\rangle\langle q2\rangle}\,\frac{1}{\langle23\rangle[23]}\,\frac{\langle 1q\rangle^3}{\langle qP\rangle\langle P1\rangle} ~=~ \frac{\omega_P(\omega_2z_{12}+\omega_3z_{13})}{\omega_2\omega_3\bz_{23}} ~~,
\label{eq:t-diagram}\\
  \begin{tikzpicture}[baseline={([yshift=-0.9ex]current bounding box.center)},scale=0.9]
        \draw[thick] (-1,0) node[above right]{$-q^{+}$} --  (1,0) node[above left]{$q^-$};
        \draw[thick] (-1,0) -- (-1.5,1.73/2) node[left]{$1^-$};
        \draw[thick] (1,0) -- (1.5,1.73/2) node[right]{$-P^+$};
        \draw[thick] (-1,0) -- (-1.5,-1.73/2) node[left]{$3^-$};
        \draw[thick] (1,0) -- (1.5,-1.73/2) node[right]{$2^-$};
\end{tikzpicture} ~=&~ \frac{\langle13\rangle^3}{\langle3q\rangle\langle q1\rangle}\,\frac{1}{\langle13\rangle[13]}\,\frac{\langle q2\rangle^3}{\langle2P\rangle\langle Pq\rangle} ~=~ \frac{\omega_P(\omega_1z_{12}+\omega_3z_{32})}{\omega_1\omega_3\bz_{13}} ~~, \label{eq:u-diagram}
\end{align}
where $\omega_P=\omega_1+\omega_2+\omega_3$.\,\footnote{In this calculation it is crucial that the particle $P$ is off-shell, following CSW rules, since otherwise the sum of the three contributions would vanish.} Together with the color factors and the propagator
\begin{equation}
    \frac{1}{s_{123}} ~=~ \frac{1}{\omega_1\omega_2\bz_{12}z_{12}+\omega_1\omega_3\bz_{13}z_{13}+\omega_2\omega_3\bz_{23}z_{23}}
\end{equation}
the splitting function can be simplified to 
\begin{equation}
    \begin{split}
        &\oh^{a,-}_1(z_1,\bz_1)\,\oh^{b,-}_2(z_2,\bz_2)\,\oh^{c,-}_3(z_3,\bz_3) \\
        &\qquad~\sim~ \Bigg[\,f^{abe}f^{ecd}\,\frac{\omega_P}{\omega_1\omega_2\omega_3\bz_{12}\bz_{23}} ~-~ f^{ace}f^{ebd}\,\frac{\omega_P}{\omega_1\omega_2\omega_3\bz_{13}\bz_{23}} \,\Bigg]\,\oh_P^{d,-}(z_3,\bz_3) ~~.
    \end{split}
    \label{equ:Split---}
\end{equation}
Note that (\ref{equ:Split---}) holds for any relative collinearity between 1, 2, and 3. 
To extract ${\cal C}^{a^-b^-c^-}{}_{d^-}$, we expand the above splitting function around $\bz_{23}$, and we obtain
\begin{equation}
    \begin{split}
        &\oh^{a,-}_1(z_1,\bz_1)\,\oh^{b,-}_2(z_2,\bz_2)\,\oh^{c,-}_3(z_3,\bz_3) \\
        ~\sim&~ \left\{f^{abe}f^{ecd}\,\frac{\omega_P}{\omega_1\omega_2\omega_3}\frac{1}{\bz_{23}}\left[ \frac{1}{\bz_{13}} + \frac{\bz_{23}}{\bz_{13}^2} + O(\bz^2_{23}) \right]
        ~-~ f^{ace}f^{ebd}\,\frac{\omega_P}{\omega_1\omega_2\omega_3}\frac{1}{\bz_{13}\bz_{23}}\right\}\,\oh^{d,-}_P(z_3,\bz_3) \\
        ~=&~ f^{bce}f^{aed}\,\frac{\omega_P}{\omega_1\omega_2\omega_3}\frac{1}{\bz_{13}\bz_{23}}\,\oh^{d,-}_P(z_3,\bz_3) 
        ~+~ f^{abe}f^{ecd}\,\frac{\omega_P}{\omega_1\omega_2\omega_3}\,\frac{1}{\bz^2_{13}} \,\oh^{d,-}_P(z_3,\bz_3) ~+~ O(\bz_{23})~~.
    \end{split} 
    \label{equ:---final}
\end{equation}
Since we're considering a consecutive collinear limit, the left-hand side of the above equation can also be written as 
\begin{equation}
    \begin{aligned}
    \oh^{a,-}_1(z_1,\bz_1)\,&\oh^{b,-}_2(z_2,\bz_2)\,\oh^{c,-}_3(z_3,\bz_3) \\
        ~\sim&~ \oh^{a,-}_1(z_1,\bz_1)\,\Bigg( f^{bce}\,\frac{\omega_q}{\omega_2\omega_3}\frac{1}{\bz_{23}}\,\oh^{e,-}_q(z_3,\bz_3) ~+~ :\oh^{b,-}_2\oh^{c,-}_3:(z_3,\bz_3)  \Bigg) ~~,
    \end{aligned}
    \label{equ:consecutive---}
\end{equation}
where we have plugged in (\ref{equ:O-O-toO-mom}). Applying (\ref{equ:O-O-toO-mom}) again to the first term above matches with the first term in (\ref{equ:---final}). Comparing the second terms in (\ref{equ:---final}) and (\ref{equ:consecutive---}) yields 
\begin{equation}
    \oh^{a,-}_1(z_1,\bz_1)\,:\oh^{b,-}_2\oh^{c,-}_3:(z_3,\bz_3) ~\sim~ f^{abe}f^{ecd}\,\frac{\omega_P}{\omega_1\omega_2\omega_3}\,\frac{1}{\bz_{13}^2}\,\oh^{d,-}_P(z_3,\bz_3) ~~.
    \label{equ:3ope-}
\end{equation}
Namely,
\begin{equation}\boxed{~
    {\cal C}^{a^-b^-c^-}{}_{d^-}(\omega_1,\omega_2,\omega_3) ~=~ f^{abe}f^{ecd}\,\frac{\omega_1+\omega_2+\omega_3}{\omega_1\omega_2\omega_3} ~~. ~} 
    \label{equ:C---}
\end{equation}

\paragraph{Associativity}
Above we considered $2\to 3$ first, then $1\to 3$. Now let us consider $1\to 2$ first, then $2\to 3$. We now expand the splitting function (\ref{equ:Split---}) around $\bz_{12}$ to find
\begin{equation}
    \begin{split}
        \oh^{a,-}_1(z_1,\bz_1)\,&\oh^{b,-}_2(z_2,\bz_2)\,\oh^{c,-}_3(z_3,\bz_3) \\
        ~\sim&~ \Bigg( f^{abe}\,\frac{\omega_q}{\omega_1\omega_2}\frac{1}{\bz_{12}}\,\oh^{e,-}_q(z_2,\bz_2) ~+~ :\oh^{a,-}_1\oh^{b,-}_2:(z_2,\bz_2)  \Bigg)\,\oh^{c,-}_3(z_3,\bz_3) \\
        ~\sim&~ f^{abe}f^{ecd}\,\frac{\omega_P}{\omega_1\omega_2\omega_3}\,\frac{1}{\bz_{12}\bz_{23}}\,\oh_P^{d,-}(z_3,\bz_3) ~+~ f^{ace}f^{bed}\,\frac{\omega_P}{\omega_1\omega_2\omega_3}\,\frac{1}{\bz_{23}^2}\,\oh^{d,-}_P(z_3,\bz_3) ~~.
    \end{split}
\end{equation}
We can see that associativity is satisfied. 
Moreover, by comparing with (\ref{equ:3ope-}), one can see that
\begin{equation}
    :\oh^{a,-}_1\oh^{b,-}_2:(z_2,\bz_2)\,\oh^{c,-}_3(z_3,\bz_3) ~=~ \oh^{c,-}_3(z_3,\bz_3)\,:\oh^{a,-}_1\oh^{b,-}_2:(z_2,\bz_2) ~~.
\end{equation}

\subsection{Mixed-helicity Splitting Functions}\label{sec:mixedh-split}

We now move on to the mixed-helicity cases, computing the following two splitting functions ${\rm Split}[1^{-,a}2^{+,b}3^{+,c}\to P^{-,d}]$ and ${\rm Split}[1^{-,a}2^{-,b}3^{+,c}\to P^{+,d}]$. 

\paragraph{\texorpdfstring{${\rm Split}[1^{-,a}2^{+,b}3^{+,c}\to P^{-,d}]$}{1-2+3+toP-}} 

First, note that all of the three-point vertices in the diagrams in (\ref{equ:diagrams}) are $\overline{\rm MHV}$ vertices. 
Hence the splitting function can be straightforwardly computed following the CSW prescription and taking the \textit{holomorphic} multicollinear limit.
Direct computation yields
\begin{align}
\begin{tikzpicture}[baseline={([yshift=-0.9ex]current bounding box.center)},scale=0.9]
        \draw[thick] (-1,0) node[above right]{$-q^{+}$} --  (1,0) node[above left]{$q^-$};
        \draw[thick] (-1,0) -- (-1.5,1.73/2) node[left]{$1^-$};
        \draw[thick] (1,0) -- (1.5,1.73/2) node[right]{$-P^+$};
        \draw[thick] (-1,0) -- (-1.5,-1.73/2) node[left]{$2^+$};
        \draw[thick] (1,0) -- (1.5,-1.73/2) node[right]{$3^+$};
\end{tikzpicture}  ~=&~ \frac{\omega_1}{\omega_2\omega_P}\frac{\omega_1\bz_{13}+\omega_2\bz_{23}}{z_{12}} ~~,\\
 \begin{tikzpicture}[baseline={([yshift=-0.9ex]current bounding box.center)},scale=0.9]
        \draw[thick] (0,-1+0.5) --  (0,1-0.5);
        \draw[thick] (0,-1+0.5) -- (1.73/2,-1.5+0.5) node[right]{$3^+$};
        \draw[thick] (0,0.5) -- (1.73/2,1.5-.5) node[right]{$-P^+$};
        \draw[thick] (0,-1+0.5) -- (-1.73/2,-1.5+0.5) node[left]{$2^+$};
        \draw[thick] (0,1-.5) -- (-1.73/2,1.5-.5) node[left]{$1^-$}; 
        \draw (0,0.4) node[right]{$q^+$};
        \draw (0,-0.4) node[right]{$-q^{-}$};
\end{tikzpicture} ~=&~ \frac{\omega^2_1}{\omega_2\omega_3\omega_P}\frac{\omega_2\bz_{12}+\omega_3\bz_{13}}{z_{23}} ~~,\\
 \begin{tikzpicture}[baseline={([yshift=-0.9ex]current bounding box.center)},scale=0.9]
        \draw[thick] (-1,0) node[above right]{$-q^{+}$} --  (1,0) node[above left]{$q^-$};
        \draw[thick] (-1,0) -- (-1.5,1.73/2) node[left]{$1^-$};
        \draw[thick] (1,0) -- (1.5,1.73/2) node[right]{$-P^+$};
        \draw[thick] (-1,0) -- (-1.5,-1.73/2) node[left]{$3^+$};
        \draw[thick] (1,0) -- (1.5,-1.73/2) node[right]{$2^+$};
\end{tikzpicture} ~=&~ \frac{\omega_1}{\omega_3\omega_P}\frac{\omega_1\bz_{12}+\omega_3\bz_{32}}{z_{13}} ~~.
\end{align}
Following the same procedure as in section~\ref{sec:singleh-split}, we have 
\begin{equation}
    \begin{aligned}
        \oh^{a,-}_1(z_1,\bz_1)&\,\oh^{b,+}_2(z_2,\bz_2)\,\oh^{c,+}_3(z_3,\bz_3) \\
        ~\sim&~ \oh^{a,-}_1(z_1,\bz_1)\,\Bigg( \frac{f^{bce}}{z_{23}}\frac{\omega_q}{\omega_2\omega_3}\,\oh^{e,+}_q(z_3,\bz_3)
        ~+~ :\oh^{b,+}_2\oh^{c,+}_3:(z_3,\bz_3) \Bigg)\\
        ~\sim&~  f^{bce}f^{aed}\,\frac{\omega_1}{\omega_P\omega_2\omega_3}\frac{1}{z_{13}z_{23}}\,\oh_P^{d,-}(z_3,\bz_3) ~+~ f^{abe}f^{ecd}\,\frac{1}{z_{13}^2}\frac{\omega_1}{\omega_2\omega_P\omega_3}\,\oh^{d,-}_P(z_3,\bz_3)
    \end{aligned}
    \label{equ:-++final}
\end{equation}
where we have plugged in (\ref{equ:O+O+toO+mom}) in the second line. 
The first term in (\ref{equ:-++final}) corresponds to applying (\ref{equ:O+O+toO+mom}) and (\ref{equ:O-O+toO+mom}) consecutively. Meanwhile, the second term yields
\begin{equation}
    \oh^{a,-}_1(z_1,\bz_1)\,:\oh^{b,+}_2\oh^{c,+}_3:(z_3,\bz_3) ~\sim~ f^{abe}f^{ecd}\,\frac{1}{z_{13}^2}\frac{\omega_1}{\omega_2\omega_P\omega_3}\,\oh^{d,-}_P(z_3,\bz_3)~~.
    \label{eq:3ope-++}
\end{equation}
Namely, the ${\cal C}^{a^-b^+c^+}{}_{d^-}$ coefficient in (\ref{eq:O:OO:OPE-++}) is
\begin{equation}\boxed{~
    {\cal C}^{a^-b^+c^+}{}_{d^-}(\omega_1,\omega_2,\omega_3) ~=~ f^{abe}f^{ecd}\,\frac{\omega_1}{\omega_2\omega_3\,(\omega_1+\omega_2+\omega_3)} ~~.
    ~}
\end{equation}

\paragraph{\texorpdfstring{${\rm Split}[1^{-,a}2^{-,b}3^{+,c}\to P^{+,d}]$}{1-2-3+toP+}} 
For this helicity configuration, direct computation yields 
\begin{equation}
    \begin{split}
        \begin{tikzpicture}[baseline={([yshift=-0.9ex]current bounding box.center)},scale=0.9]
        \draw[thick] (-1,0) node[above right]{$-q^{+}$} --  (1,0) node[above left]{$q^-$};
        \draw[thick] (-1,0) -- (-1.5,1.73/2) node[left]{$1^-$};
        \draw[thick] (1,0) -- (1.5,1.73/2) node[right]{$-P^-$};
        \draw[thick] (-1,0) -- (-1.5,-1.73/2) node[left]{$2^-$};
        \draw[thick] (1,0) -- (1.5,-1.73/2) node[right]{$3^+$};
\end{tikzpicture} ~=&~ \frac{\omega_3^2}{\omega_1\omega_2\omega_P}\,\frac{(\omega_1z_{13}+\omega_2z_{23})}{\bz_{12}} ~~,\\
\begin{tikzpicture}[baseline={([yshift=-0.9ex]current bounding box.center)},scale=0.9]
        \draw[thick] (-1,0) node[above right]{$-q^{-}$} --  (1,0) node[above left]{$q^+$};
        \draw[thick] (-1,0) -- (-1.5,1.73/2) node[left]{$1^-$};
        \draw[thick] (1,0) -- (1.5,1.73/2) node[right]{$-P^-$};
        \draw[thick] (-1,0) -- (-1.5,-1.73/2) node[left]{$3^+$};
        \draw[thick] (1,0) -- (1.5,-1.73/2) node[right]{$2^-$};
\end{tikzpicture} ~=&~ \frac{\omega_3}{\omega_1\omega_P}\,\frac{(\omega_1z_{12}+\omega_3z_{32})}{\bz_{13}} ~~,\\
\begin{tikzpicture}[baseline={([yshift=-0.9ex]current bounding box.center)},scale=0.9]
        \draw[thick] (0,-1+0.5) --  (0,1-0.5);
        \draw[thick] (0,-1+0.5) -- (1.73/2,-1.5+0.5) node[right]{$3^+$};
        \draw[thick] (0,0.5) -- (1.73/2,1.5-.5) node[right]{$-P^-$};
        \draw[thick] (0,-1+0.5) -- (-1.73/2,-1.5+0.5) node[left]{$2^-$};
        \draw[thick] (0,1-.5) -- (-1.73/2,1.5-.5) node[left]{$1^-$}; 
        \draw (0,0.4) node[right]{$q^+$};
        \draw (0,-0.4) node[right]{$-q^{-}$};
\end{tikzpicture} ~=&~ \frac{\omega_3}{\omega_2\omega_P}\,\frac{(\omega_2z_{12}+\omega_3z_{13})}{\bz_{23}} ~~.
    \end{split}
\end{equation}
After expanding in $\bz_{23}$, we obtain
\begin{equation}
    \begin{aligned}
        \oh^{a,-}_1(z_1,\bz_1)&\,\oh^{b,-}_2(z_2,\bz_2)\,\oh^{c,+}_3(z_3,\bz_3) \\
        ~\sim&~ \oh^{a,-}_1(z_1,\bz_1)\,\Bigg( f^{bce}\,\frac{\omega_3}{\omega_2\omega_q}\,\frac{1}{\bz_{23}}\,\oh_q^{e,+}(z_3,\bz_3)
        ~+~ :\oh^{b,-}_2\oh^{c,+}_3:(z_3,\bz_3) \Bigg)\\
        ~\sim&~  f^{bce}f^{aed}\,\frac{\omega_3}{\omega_1\omega_2\omega_P}\,\frac{1}{\bz_{13}\bz_{23}}\,\oh_P^{d,+}(z_3,\bz_3) ~+~ f^{abe}f^{ecd}\,\frac{1}{\bz_{13}^2}\,\frac{\omega_3}{\omega_1\omega_2\omega_P}\,\oh^{d,+}_P(z_3,\bz_3)
    \end{aligned}
    \label{equ:--+final}
\end{equation}
where we have plugged in (\ref{equ:O-O+toO+mom}) in the second line. The first term in (\ref{equ:--+final}) corresponds to applying (\ref{equ:O-O+toO+mom}) twice, while the second term yields
\begin{equation}
    \oh^{a,-}_1(z_1,\bz_1)\,:\oh^{b,-}_2\oh^{c,+}_3:(z_3,\bz_3) ~\sim~ f^{abe}f^{ecd}\,\frac{1}{\bz_{13}^2}\,\frac{\omega_3}{\omega_1\omega_2\omega_P}\,\oh^{d,+}_P(z_3,\bz_3)~~.
    \label{eq:3ope--+}
\end{equation}
Namely the ${\cal C}^{a^-b^-c^+}{}_{d^+}$ coefficient in (\ref{eq:O:OO:OPE--+}) is
\begin{equation}\boxed{~
    {\cal C}^{a^-b^-c^+}{}_{d^+}(\omega_1,\omega_2,\omega_3) ~=~ f^{abe}f^{ecd}\,\frac{\omega_3}{\omega_1\omega_2\,(\omega_1+\omega_2+\omega_3)} ~~.
    ~}
\end{equation}

\subsection{Celestial OPE Coefficients}\label{sec:Mellin-transf}

To extract the celestial OPE coefficients, we now implement Mellin transforms of $\oh_1$, $\oh_2$, and $\oh_3$ on both sides of equations (\ref{equ:3ope-}), (\ref{eq:3ope-++}), and (\ref{eq:3ope--+}). Since their $\omega$ dependence is similar, we consider the generic case below. 

First, the measure of the Mellin transforms can be rewritten as follows
\begin{equation}
    \begin{split}
        &\int_0^{+\infty}d\omega_1\,\omega_1^{\D_1-1}\int_0^{+\infty}d\omega_2\,\omega_2^{\D_2-1}\int_0^{+\infty}d\omega_3\,\omega_3^{\D_3-1} \\
        ~=&~ \int_0^{+\infty}d\omega_P\,\omega_P^{\D_1+\D_2+\D_3-1}\,\int_0^1\frac{d\s_1}{\s_1}\s_1^{\D_1}\int_0^1\frac{d\s_2}{\s_2}\s_2^{\D_2}\int_0^1\frac{d\s_3}{\s_3}\s_3^{\D_3}\,\d(1-\s_1-\s_2-\s_3)
    \end{split}
\end{equation}
where
\begin{equation}
    \omega_P~=~ \omega_1+\omega_2+\omega_3 ~~,~~ \s_1 ~=~ \frac{\omega_1}{\omega_P}~~,~~ \s_2 ~=~ \frac{\omega_2}{\omega_P}~~,~~ \s_3 ~=~ \frac{\omega_3}{\omega_P}~~,~~ {\rm Jacobian}~=~ \omega_P^2~~.
\end{equation}
Consider a general integrand taking the form $\omega^{n_1}_1\omega^{n_2}_2\omega^{n_3}_3\omega^{n_P}_P$, where the $n$'s are arbitrary numbers. 
We then have
\begin{equation}
    \begin{split}
        &\left(\prod_{i=1}^3\,\int_0^{+\infty}d\omega_i\,\omega_i^{\D_i-1}\right)\,\omega^{n_1}_1\omega^{n_2}_2\omega^{n_3}_3\omega^{n_P}_P ~=~ \int_0^{+\infty}d\omega_P\,\omega_P^{\D_1+\D_2+\D_3-1+n_P+n_1+n_2+n_3} \\
        &\qquad~\times~\int_0^1\frac{d\s_1}{\s_1}\s_1^{\D_1}\int_0^1\frac{d\s_2}{\s_2}\s_2^{\D_2}\int_0^1\frac{d\s_3}{\s_3}\s_3^{\D_3}\,\d(1-\s_1-\s_2-\s_3)\,\s^{n_1}_1\s^{n_2}_2\s^{n_3}_3 ~~.
    \end{split}
\end{equation}
The integral over the $\s$'s gives the OPE coefficient, which reads
\begin{equation}
    B(\D_1+n_1,\D_2+n_2,\D_3+n_3) ~=~ \frac{\Gamma(\D_1+n_1)\Gamma(\D_2+n_2)\Gamma(\D_3+n_3)}{\Gamma(\D_1+\D_2+\D_3+n_1+n_2+n_3)} ~~.
\end{equation}
On the RHS of (\ref{equ:3ope-}), (\ref{eq:3ope-++}), and (\ref{eq:3ope--+}), note that $\oh^d_P$ is in the momentum basis with energy $\omega_P$, which can be related to the boost basis via the inverse Mellin transform. 
Then the $\omega_P$ integral turns into $\oh_{\D_1+\D_2+\D_3+n_P+n_1+n_2+n_3}^d(z_3,\bz_3)$. Namely, 
\begin{equation}
    \begin{split}
    &\int_0^{+\infty}d\omega_P\,\omega_P^{\D_1+\D_2+\D_3-1+n_P+n_1+n_2+n_3}\,\frac{1}{2\pi i}\,\int d\Delta_P\,\omega_P^{-\D_P}\,\oh_{\D_P,J_P}^{d} \\
    ~=&~ \int d\Delta_P\,\d(\D_1+\D_2+\D_3+n_P+n_1+n_2+n_3-\D_P)\,\oh_{\D_P,J_P}^{d}
    \end{split}
\end{equation}
where we have used the identity
\begin{equation}
    \int_0^{+\infty}\,d\omega\, \omega^{\D-1} ~=~ 2\pi i\,\d(\D)~~.
\end{equation}
Therefore, in the boost basis, the ${\cal C}$ coefficients in (\ref{eq:O:OO:OPE}) become\,\footnote{Note that in the celestial OPEs of the form (\ref{eq:O:OO:OPE}), there is implicitly a $\D_P$ integral in the index contraction. Schematically, the OPE takes the form 
$\oh_1\oh_2~\sim~ \sum_{J_P}\int d\D_P C_{\D_1,\D_2}{}^{\D_P}\oh_{\D_P,J_P}$ and we have fixed the spin $J_P$. A similar statement holds in momentum space where $\D$ is traded with $\omega$.}
\begin{equation}\boxed{~
\begin{aligned}
    {\cal C}^{a^-b^-c^-}{}_{d^-} ~=&~ f^{abe}f^{ecd}\,\frac{\Gamma(\D_1-1)\Gamma(\D_2-1)\Gamma(\D_3-1)}{\Gamma(\D_1+\D_2+\D_3-3)}\,\d(\D_1+\D_2+\D_3-2-\D_P) ~~,\\
    {\cal C}^{a^-b^+c^+}{}_{d^-} ~=&~ f^{abe}f^{ecd}\,\frac{\Gamma(\D_1+1)\Gamma(\D_2-1)\Gamma(\D_3-1)}{\Gamma(\D_1+\D_2+\D_3-1)}\,\d(\D_1+\D_2+\D_3-2-\D_P) ~~, \\
    {\cal C}^{a^-b^-c^+}{}_{d^+} ~=&~ f^{abe}f^{ecd}\,\frac{\Gamma(\D_1-1)\Gamma(\D_2-1)\Gamma(\D_3+1)}{\Gamma(\D_1+\D_2+\D_3-1)}\,\d(\D_1+\D_2+\D_3-2-\D_P) ~~.
\end{aligned}~}
\label{equ:celestial-OPE-coeff-single}
\end{equation}

\section{Two-particle Exchanges}\label{sec:2ptexchange}

So far we have computed single-particle exchanges in the OPE between a single-particle and two-particle operator. These single-particle contributions are captured by 4-particle factorization channels, while multiparticle exchanges do \textit{not} appear in this analysis. 
This section is devoted to showing that multiparticle exchanges can be extracted from evaluating the full amplitude. 
For simplicity, we consider the single-helicity sector. Namely, all operators that appear in the OPE have the same helicity, which can be effectively described by MHV or $\overline{\rm MHV}$ amplitudes.\,\footnote{As mentioned above, one needs to consider ${\rm N}^k{\rm MHV}$ amplitudes to get other helicity configurations in the OPE.}

In this section, we consider a generic $n$-point tree-level $\overline{\rm MHV}$ gluon amplitude. Without loss of generality, we choose the helicity configuration where particles $n-1$ and $n$ have positive helicity. 
To extract the OPE (\ref{eq:O:OO:OPE---}), we expand $\bz_2$ around $\bz_3$ first. From (\ref{equ:O-O-toO-mom}) we get
\begin{equation}
   \begin{split}
       \langle \oh_1^{a_1,-}\,\oh_2^{a_2,-}\,\oh_3^{a_3,-}\,\oh_4^{a_4,-}\cdots\oh_{n-1}^{a_{n-1},+}\oh_n^{a_n,+}\rangle ~=&~ \frac{f^{a_2a_3a_q}}{\bz_{23}}\,\frac{\omega_q}{\omega_2\omega_3}\, \langle \oh_1^{a_1,-}\,\oh_q^{a_q,-}(3)\,\oh_4^{a_4,+}\cdots\oh_n^{a_n,+}\rangle \\
       ~+~ (\bz_{23}^0)\,\langle \oh_1^{a_1,-}\,&:\oh_2^{a_2,-}\oh_3^{a_3,-}:(3)\,\oh_4^{a_4,+}\cdots\oh_n^{a_n,+}\rangle ~+~ O(\bz_{23})~~,
   \end{split} 
\end{equation}
where $(3)$ is short for $(z_3,\bz_3)$ and, to lighten the expression, we omit the positions on the celestial sphere when they are unambiguous, i.e. the $\oh_i$ operator is inserted at $(z_i,\bz_i)$ with energy $\omega_i$ unless specifically labeled. 
In what follows we will only keep the leading term (at order $1/\bz_{23}$) and subleading term (at order $\bz_{23}^0$). 
Next, we expand $\bz_1$ around $\bz_3$ and only keep the singular terms in $\bz_{13}$. 
As we will see shortly, the leading term ($1/\bz_{23}$) correctly reproduces the consecutive single-particle OPE, i.e. the first term in (\ref{equ:---final}). The subleading ($\bz_{23}^0$) terms take the following form after our $\bz_{13}$ expansion 
\begin{equation}
    \begin{split}
        \langle \oh_1^{a_1,-}\,:\oh_2^{a_2,-}&\oh_3^{a_3,-}:(3)\,\oh_4^{a_4,-}\cdots\oh_n^{a_n,+}\rangle = \frac{{\cal C}^{a^-_1,a^-_2,a^-_3}{}_{a^-_P}(\omega_1,\omega_2,\omega_3)}{\bz_{13}^2}\,\langle \oh_P^{a_P,-}(3)\oh_4^{a_4,-}\cdots\oh_n^{a_n,+}\rangle \\
        &~+~ \frac{{\cal D}^{a^-_1,a^-_2,a^-_3}{}_{a^-_q,a^-_l}(\omega_1,\omega_2,\omega_3)}{\bz_{13}}\,\langle \,:\oh_q^{a_q,-}\oh^{a_l,-}_l:(3)\,\oh_4^{a_4,-}\cdots\oh_n^{a_n,+}\rangle ~+~ O(\bz_{13})~~.
    \end{split}
    \label{equ:npt-ansatz}
\end{equation}
Note that in the RHS, the lower-point amplitudes depend on $\omega_P$, $\omega_q$, and $\omega_l$. As such, similar to the celestial OPE, there are implicitly $\omega$-integrals in the index contractions.

\subsection{Evaluation of the Full Gluon Amplitude}

Adopting the DDM basis~\cite{DelDuca:1999rs}, the full $n$-point tree-level gluon amplitude reads
\begin{equation}
    \begin{split}
        A_n^{\rm full,tree} ~=&~ \sum_{\s\in S_{n-2}} f^{a_1a_{\s_2}b_1}f^{b_1a_{\s_3}b_2}\cdots f^{b_{n-3}a_{\s_{n-1}}a_n}\,A[1\s_2\s_3\s_4\cdots\s_{n-1}n] ~~,
    \end{split}
    \label{equ:DDM}
\end{equation}
where $\s$ is summed over permutations of $\{2,\cdots,n-1\}$. 
For each color-ordered amplitude $A[1\s_2\s_3\s_4\cdots\s_{n-1}n]$, we can simply plug in the Parke–Taylor formula~\cite{PhysRevLett.56.2459,Elvang:2013cua}. 
Since we are interested in gluons 1, 2, and 3 going collinear consecutively, we further expand the $n$-point amplitude as follows 
\begin{equation}
    \begin{split}
        A_n^{\rm full,tree} &~=~ 
        \left[\,\sum_{\s\in S_{n-4}} f^{a_1a_{2}b_1}f^{b_1a_{3}b_2}f^{b_2a_{\s_4}b_3}\cdots f^{b_{n-3}a_{\s_{n-1}}a_n}\,A[123\s_4\cdots\s_{n-1}n] ~+~ (2\leftrightarrow3)\,\right]
        \\
        &~+~ 
        \left[\,\sum_{\s\in S_{n-4}}\sum_{i=4}^{n-1} f^{a_1a_{2}b_1}f^{b_1a_{i}b_2}f^{b_2a_{\s_4}b_3}\cdots f^{b_{n-3}a_{\s_{n-1}}a_n}\,A[12i\s_4\cdots\s_{n-1}n] ~+~ (2\leftrightarrow3)\,\right] 
        \\
        &~+~ \sum_{i=4}^{n-1} \sum_{j=2,j\ne i}^{n-1}\,\sum_{\s\in S_{n-4}} f^{a_1a_{i}b_1}f^{b_1a_{j}b_2}f^{b_2a_{\s_4}b_3}\cdots f^{b_{n-3}a_{\s_{n-1}}a_n}\,A[1ij\s_4\cdots\s_{n-1}n] ~~.
    \end{split}
    \label{equ:npt-full}
\end{equation}
Note that the full amplitude also contains momentum-conserving delta functions that we've omitted in (\ref{equ:DDM}) and (\ref{equ:npt-full}). It turns out that since we are focusing on primary operators in the OPE, 
one can strip off the momentum-conserving delta functions and consider the expansion of the stripped amplitude (\ref{equ:npt-full}).\,\footnote{Note that in the $z,\bz$-expansions of the delta functions, the leading terms are evaluated at $z_{23}=\bz_{23}=z_{13}=\bz_{13}=0$, which turns into the momentum-conserving delta functions for the lower point amplitudes. The subleading terms contain $z,\bz$ derivatives, and after integration by part, they turn into descendants. We don't consider descendants in this paper. }
For the $\bz$-expansions, let's use the first term in (\ref{equ:npt-full}) as an example. Plugging in the Parke-Taylor formula gives 
\begin{equation}
    A[123\s_4\cdots\s_{n-1}n] ~=~ \frac{[n-1,n]^4}{[12][23][3\s_4]\cdots [n1]} ~=~ \frac{\omega_{n-1}\omega_n}{\omega_1\omega_2\omega_3\cdots\omega_{n-2}}\,\frac{\bz_{n-1,n}^4}{\bz_{12}\bz_{23}\bz_{3\s_4}\cdots\bz_{n1}} ~~.
\end{equation}
Expanding $\bz_2$ around $\bz_3$ first, and then $\bz_1$ around $\bz_3$ yields 
\begin{equation}
    \begin{split}
        &A[123\s_4\cdots\s_{n-1}n] \\
        ~=&~ \frac{\omega_{n-1}\omega_n}{\omega_1\omega_2\omega_3\cdots\omega_{n-2}}\,\frac{\bz_{n-1,n}^4}{\bz_{23}\bz_{3\s_4}\cdots\bz_{\s_{n-1},n}}\left( \frac{1}{\bz_{13}} + \frac{\bz_{23}}{\bz^2_{13}} + \cdots\right)\left( \frac{1}{\bz_{n3}} + \frac{\bz_{13}}{\bz^2_{n3}} + \cdots \right) \\
        ~=&~ \frac{\omega_{n-1}\omega_n}{\omega_1\omega_2\omega_3\cdots\omega_{n-2}}\,\frac{\bz_{n-1,n}^4}{\bz_{3\s_4}\cdots\bz_{\s_{n-1},n}}\,\left( \frac{1}{\bz_{13}\bz_{23}}\,\frac{1}{\bz_{n3}}+ \frac{1}{\bz_{13}^2}\,\frac{1}{\bz_{n3}} + \frac{1}{\bz_{13}}\,\frac{1}{\bz^2_{n3}} + O(\bz^0_{13},\bz_{23}) \right) ~~.
    \end{split}
    \label{equ:A1subleading}
\end{equation}
All the other color-ordered amplitudes in (\ref{equ:npt-full}) can be expanded in a similar manner. The leading terms ($1/\bz_{23}$) and subleading terms ($\bz_{23}^0$) in the final answer are summarized below.

\paragraph{Leading terms}
Note that only the two color-ordered amplitudes $A[123\s_4\cdots\s_{n-1}n]$ and $A[132\s_4\cdots\s_{n-1}n]$ have both $1/\bz_{23}$ and $1/\bz_{13}$ singularities. The leading terms read
\begin{equation}
  \begin{split}
       \langle \oh_1^{a_1,-}\,\oh_2^{a_2,-}\,\oh_3^{a_3,-}\,\oh_4^{a_4,-}\cdots\oh_{n-1}^{a_{n-1},+}\oh_n^{a_n,+}\rangle 
   ~\supset~ \Big(f^{a_1a_2b}f^{ba_3a_P} - f^{a_1a_3b}f^{ba_2a_P}\Big)\,\frac{\omega_P}{\omega_1\omega_2\omega_3} &\\
    ~\times~\frac{1}{\bz_{13}\bz_{23}}\, \sum_{\s\in S_{n-4}} f^{a_Pa_{\s_4}b_3}\cdots f^{b_{n-3}a_{\s_{n-1}}a_n}\,\frac{\omega_{n-1}\omega_n}{\omega_P\omega_4\cdots\omega_{n-2}}\frac{\bz_{n-1,n}^4}{\bz_{3\s_4}\bz_{\s_4\s_5}\cdots\bz_{\s_{n-1}n}\bz_{n3}} &\\
   ~=~ f^{a_2a_3b}f^{a_1ba_P}\,\frac{\omega_P}{\omega_1\omega_2\omega_3}\,\frac{1}{\bz_{13}\bz_{23}}\,\langle\oh_P^{a_P,-}(3)\oh_4^{a_4,-}\cdots\oh_n^{a_n,+}\rangle ~~, &
  \end{split}
  \label{equ:n-pt-leading}
\end{equation}
where $\oh_P^{a_P,-}(3)$ is an exchanged gluon $P^-$ located at $(z_3,\bz_3)$ with energy $\omega_P=\omega_1+\omega_2+\omega_3$ 
and the remaining $(n-2)$-point amplitude takes the following form in the DDM basis
\begin{equation}
    \langle\oh_P^{a_P,-}(3)\oh_4^{a_4,-}\cdots\oh_n^{a_n,+}\rangle ~=~ \sum_{\s\in S_{n-4}} f^{a_Pa_{\s_4}b_3}\cdots f^{b_{n-3}a_{\s_{n-1}}a_n}\,A[P(3)\s_4\cdots\s_{n-1}n] ~~.
    \label{equ:n-2pt}
\end{equation}
Given (\ref{equ:n-pt-leading}), we extract the following contribution to the OPE in the momentum basis
\begin{equation}
    \oh_1^{a_1,-}(z_1,\bz_1)\,\oh_2^{a_2,-}(z_2,\bz_2)\,\oh_3^{a_3,-}(z_3,\bz_3) ~\sim~ f^{a_2a_3b}f^{a_1ba_P}\,\frac{\omega_P}{\omega_1\omega_2\omega_3}\,\frac{1}{\bz_{13}\bz_{23}}\, \oh_P^{a_P,-}(z_3,\bz_3) ~~,
\end{equation}
which exactly matches with the first term in (\ref{equ:---final}).

\paragraph{Subleading terms}
For the subleading terms, we first note that since the terms in the last line of (\ref{equ:npt-full}) are regular in $\bz_{13}$, they will not contribute to the sector of the OPEs that we are discussing here. Evaluating the terms in the first and second lines of (\ref{equ:npt-full}) gives
\begin{equation}
    \begin{split}
        & \langle \oh_1^{a_1,-}\,:\oh_2^{a_2,-}\oh_3^{a_3,-}:(3)\,\oh_4^{a_4,-}\cdots\oh_n^{a_n,+}\rangle \\
         &~=~ \frac{1}{\bz_{13}^2}\frac{\omega_P}{\omega_1\omega_2\omega_3}\,\sum_{\s\in S_{n-4}} f^{a_1a_{2}b_1}f^{b_1a_{3}b_2}f^{b_2a_{\s_4}b_3}\cdots f^{b_{n-3}a_{\s_{n-1}}a_n}\,A[P(3)\s_4\cdots\s_{n-1}n] \\
         &~+~ \frac{1}{\bz_{13}}\frac{\omega_{n-1}\omega_n}{\omega_1\omega_2\cdots\omega_{n-2}}\,\Bigg[\sum_{\s\in S_{n-4}} f^{a_1a_{2}b_1}f^{b_1a_{3}b_2}f^{b_2a_{\s_4}b_3}\cdots f^{b_{n-3}a_{\s_{n-1}}a_n}\,\frac{\bz_{n-1,n}^4}{\bz_{3\s_4}\bz_{\s_4\s_5}\cdots\bz_{\s_{n-1}n}\bz_{n3}^2} \\
         &\qquad~+~ \sum_{\s\in S_{n-4}}\sum_{k=4}^{n-1} f^{a_1a_{2}b_1}f^{b_1a_{k}b_2}f^{b_2a_{\s_4}b_3}\cdots f^{b_{n-3}a_{\s_{n-1}}a_n}\,\frac{\bz_{n-1,n}^4}{\bz_{3k}\bz_{k\s_4}\cdots\bz_{\s_{n-1}n}\bz_{n3}} \\
         &\qquad~+~ \sum_{\s\in S_{n-4}} f^{a_1a_{3}b_1}f^{b_1a_{2}b_2}f^{b_2a_{\s_4}b_3}\cdots f^{b_{n-3}a_{\s_{n-1}}a_n}\,\frac{\bz_{n-1,n}^4}{\bz^2_{3\s_4}\bz_{\s_4\s_5}\cdots\bz_{\s_{n-1}n}\bz_{n3}} \\
         &\qquad~+~ \sum_{\s\in S_{n-4}}\sum_{k=4}^{n-1} f^{a_1a_{3}b_1}f^{b_1a_{k}b_2}f^{b_2a_{\s_4}b_3}\cdots f^{b_{n-3}a_{\s_{n-1}}a_n}\,\frac{\bz_{n-1,n}^4}{\bz_{3k}\bz_{k\s_4}\cdots\bz_{\s_{n-1}n}\bz_{n3}} 
         \Bigg] ~+~ O(\bz_{13}^0) ~~.
    \end{split}
    \label{equ:nptO2O3}
\end{equation}
Given (\ref{equ:n-2pt}), the first term above becomes
\begin{equation} 
 \frac{f^{a_1a_2b}f^{ba_3a_P}}{\bz_{13}^2}\frac{\omega_P}{\omega_1\omega_2\omega_3}\,\langle \oh_P^{a_P,-}(3)\oh_4^{a_4,-}\cdots\oh_n^{a_n,+}\rangle ~~,
\label{equ:Cextraction}
\end{equation}
which turns into the first term in the RHS of (\ref{equ:npt-ansatz}). The ${\cal C}$ coefficient then reads
\begin{equation}
    {\cal C}^{a^-_1,a^-_2,a^-_3}{}_{a^-_P} ~=~  f^{a_1a_2b}f^{ba_3a_P}\frac{\omega_P}{\omega_1\omega_2\omega_3} ~~,
\end{equation}
which matches (\ref{equ:C---}). 
The terms at order $1/\bz_{13}$ in (\ref{equ:nptO2O3}) can be reorganized in terms of the subleading terms of the $(n-1)$-point amplitude, from which we extract the ${\cal D}$ coefficient in (\ref{equ:npt-ansatz}). Indeed, the $\omega$-factor can be rearranged in the following way
\begin{equation}
    \frac{\omega_{n-1}\omega_n}{\omega_1\omega_2\cdots\omega_{n-2}} ~=~ \int_0^{\omega_P}\,d\omega_q\,f(\omega_q;\omega_1,\omega_2,\omega_3)\,\frac{\omega_q\omega_l}{\omega_1\omega_2\omega_3}\,\frac{\omega_{n-1}\omega_n}{\omega_q\omega_l\,\omega_4\cdots\omega_{n-2}}
    \label{equ:factorize-omega}
\end{equation}
where $f(\omega_q;\omega_1,\omega_2,\omega_3)$ is an undetermined function satisfying 
\begin{equation}
    \int_0^{\omega_P}\,d\omega_q\,f(\omega_q;\omega_1,\omega_2,\omega_3) ~=~ 1 
    ~~\overset{\omega_q=t\omega_P}{\longrightarrow}~~ 
    \int_0^{1}\,\omega_P\,dt\,f(t\omega_P;\omega_1,\omega_2,\omega_3) ~=~ 1 ~~.
    \label{equ:f(oq)-cond}
\end{equation}
Note that the momentum conservation yields $\omega_1+\omega_2+\omega_3=\omega_P=\omega_q+\omega_l$. Namely, only the total energy is constrained and the function $f$ essentially plays the role weighting each pair of $(\omega_q,\omega_l)$. 
Moreover, the scale symmetry for energy requires $f(t\omega_P;\omega_i)$ is a homogeneous function of degree $-1$, namely,
\begin{equation}
    f(\lambda t\omega_P;\lambda\omega_1,\lambda\omega_2,\lambda\omega_3) ~=~ \lambda^{-1}\,f( t\omega_P;\omega_1,\omega_2,\omega_3) ~~.
\end{equation}
Therefore, the function $f(t\omega_P;\omega_1,\omega_2,\omega_3)$ can be expressed as 
\begin{equation}
    f(t\omega_P;\omega_1,\omega_2,\omega_3) ~:=~ \frac{1}{\omega_P}\,\tilde{f}\left(t;\frac{\omega_1}{\omega_P},\frac{\omega_2}{\omega_P},\frac{\omega_3}{\omega_P}\right)~~,
\end{equation}
where now the function $\tilde{f}$ is an arbitrary function satisfying
\begin{equation}
   \int_0^{1}\,dt\, \tilde{f}\left(t;\frac{\omega_1}{\omega_P},\frac{\omega_2}{\omega_P},\frac{\omega_3}{\omega_P}\right)  ~=~ 1 ~~.
   \label{equ:ft-cond}
\end{equation}

Together with the $\bz$-factors, the order $1/\bz_{13}$ terms in (\ref{equ:nptO2O3}) become
\begin{equation}
    \begin{split}
       &~~ \frac{1}{\bz_{13}}\,\int_0^{1}dt\,\tilde{f}\left(t;\frac{\omega_1}{\omega_P},\frac{\omega_2}{\omega_P},\frac{\omega_3}{\omega_P}\right)\,t(1-t)\,\frac{\omega_P^2}{\omega_1\omega_2\omega_3}\\
       &~\times~\Bigg[\sum_{\s\in S_{n-4}} f^{a_1a_{2}b_1}f^{b_1a_{3}b_2}
       f^{b_2a_{\s_4}b_3}
       \cdots f^{b_{n-3}a_{\s_{n-1}}a_n}\,A[q(2)l(3)\s_4\cdots\s_{n-1}n]|_{\bz_2=\bz_3} \\
         &\qquad~+~ \sum_{\s\in S_{n-4}}\sum_{k=4}^{n-1} f^{a_1a_{2}b_1}f^{b_1a_{k}b_2}f^{b_2a_{\s_4}b_3}\cdots f^{b_{n-3}a_{\s_{n-1}}a_n}\,A[q(2)k\s_4\cdots l(3)\cdots\s_{n-1}n]|_{\bz_2=\bz_3} \\
         &\qquad\qquad\qquad\qquad\qquad\qquad\qquad\qquad\qquad\qquad\qquad\qquad ~+~ \Big( q(2)\leftrightarrow l(3)\Big)  \Bigg] ~+~ O(\bz_{13}^0) ~~,
    \end{split}
    \label{equ:two-pt-exchangelhs}
\end{equation}
where $A[\cdots]|_{\bz_2=\bz_3}$ means the coefficients at order $\bz_{23}^0$ in the $\bz_{23}$-expansion for this color-ordered amplitude.  
Note that the full $(n-1)$-point amplitude has the following two equivalent expressions in the DDM basis 
\begin{equation}
    \begin{split}
         \langle \,\oh_q^{a_q,-}(2)&\oh^{a_l,-}_l(3)\,\oh_4^{a_4,-}\cdots\oh_n^{a_n,+}\rangle \\
        ~=&~ \sum_{\s\in S_{n-4}} f^{a_qa_{l}b_2}f^{b_2a_{\s_4}b_3}\cdots f^{b_{n-3}a_{\s_{n-1}}a_n}\,A[q(2)l(3)\s_4\cdots\s_{n-1}n]  \\
        &~+~ \sum_{\s\in S_{n-4}}\sum_{k=4}^{n-1}\,f^{a_qa_{k}b_2}f^{b_2a_{\s_4}b_3}\cdots f^{b_{n-3}a_{\s_{n-1}}a_n}\,A[q(2)k\s_4\cdots l(3)\cdots\s_{n-1}n]  \\
        ~=&~ \text{the above expression with } \Big( q(2)\leftrightarrow l(3)\Big)  ~~.
    \end{split}
    \label{equ:n-1-full}
\end{equation} 
Therefore, (\ref{equ:two-pt-exchangelhs}) becomes
\begin{equation}
    \begin{split}
        &\frac{1}{\bz_{13}}\,\int_0^{1}dt\,\tilde{f}\left(t;\frac{\omega_1}{\omega_P},\frac{\omega_2}{\omega_P},\frac{\omega_3}{\omega_P}\right)\,t(1-t)\,\frac{\omega_P^2}{\omega_1\omega_2\omega_3} \\
        &\qquad \Big(f^{a_1a_2a_q}\,\d^{a_3,a_l} + f^{a_1a_3a_l}\,\d^{a_2,a_q} \Big)\langle \,:\oh_q^{a_q,-}\oh^{a_l,-}_l:(3)\,\oh_4^{a_4,-}\cdots\oh_n^{a_n,+}\rangle~~.
    \end{split}
\end{equation}
Finally, together with (\ref{equ:Cextraction}), we see that subleading terms indeed take the form of (\ref{equ:npt-ansatz}).
The ${\cal D}$ coefficient reads\,\footnote{Keep in mind that $\oh_q$ has energy $\omega_q=\omega_Pt$ and $\oh_l$ has energy $\omega_l=\omega_P(1-t)$.}
\begin{equation}\boxed{~
\begin{aligned}
   &{\cal D}^{a^-_1,a^-_2,a^-_3}{}_{a^-_q,a^-_l}(\omega_1,\omega_2,\omega_3) \\
   ~=&~  \Big(f^{a_1a_2a_q}\,\d^{a_3,a_l} + f^{a_1a_3a_l}\,\d^{a_2,a_q} \Big)\,\int_0^{1}dt\,\tilde{f}\left(t;\frac{\omega_1}{\omega_P},\frac{\omega_2}{\omega_P},\frac{\omega_3}{\omega_P}\right)\,t(1-t)\,\frac{\omega_P^2}{\omega_1\omega_2\omega_3}~~.
   \end{aligned}~}
\end{equation}

\subsection{Celestial OPE involving Two-particle Exchanges}\label{sec:celestialOPE-2pt} 

The above computation yields the following OPE in the momentum basis
\begin{equation}
    \begin{split}
        &\oh_1^{a_1,-}(z_1,\bz_1)\,:\oh_2^{a_2,-}\oh_3^{a_3,-}:(z_3,\bz_3) ~\sim~ \frac{f^{a_1a_2b}f^{ba_3a_P}}{\bz_{13}^2}\, \frac{\omega_P}{\omega_1\omega_2\omega_3}\,\oh_P^{a_P,-}(z_3,\bz_3) \\
        &\resizebox{0.9\textwidth}{!}{$+\scalemath{1.0}{\frac{\big(f^{a_1a_2a_q}\,\d^{a_3,a_l} + f^{a_1a_3a_l}\,\d^{a_2,a_q} \big)}{\bz_{13}}}\,\scalemath{1.0}{\int_0^{1}dt\tilde{f}\left(t;\frac{\omega_1}{\omega_P},\frac{\omega_2}{\omega_P},\frac{\omega_3}{\omega_P}\right)\frac{t(1-t)\omega_P^2}{\omega_1\omega_2\omega_3}}\,:\oh_q^{a_q,-}\oh^{a_l,-}_l:(z_3,\bz_3) ~~.$} 
    \end{split}
    \label{equ:O:OO:-OPE}
\end{equation}
By implementing (inverse) Mellin transforms, we obtain the celestial OPE coefficients. 
Similar to the computation in section~\ref{sec:Mellin-transf}, in the all-minus helicity sector, the celestial OPE between a single-particle operator and a two-operator operator takes the following form
\begin{equation}
    \begin{split}
        &\oh_{\D_1,-}^{a_1}(z_1,\bz_1)\,:\oh_{\D_2,-}^{a_2}\oh_{\D_3,-}^{a_3}:(z_3,\bz_3) ~\sim~ \int d\D_P\,\frac{{\cal C}^{a_1^-a_2^-a_3^-}{}_{a_P^-}(\D_1,\D_2,\D_3,\D_P)}{\bz_{13}^2}\,\oh_{\D_P,-}^{a_P}(z_3,\bz_3) \\
        &~+~ \int d\D_q\,\int d\D_l\,\frac{{\cal D}^{a_1^-a_2^-a_3^-}{}_{a_q^-a_l^-}(\D_1,\D_2,\D_3,\D_q,\D_l)}{\bz_{13}} \,:\oh^{a_q}_{\D_q,-}\oh^{a_l}_{\D_l,-}:(z_3,\bz_3)
    \end{split}
    \label{eq:celestialO:OO:}
\end{equation}
where the first term has already been evaluated in (\ref{equ:celestial-OPE-coeff-single}) as
\begin{equation}
    {\cal C}^{a_1^-a_2^-a_3^-}{}_{a_P^-}(\D_1,\D_2,\D_3,\D_P) ~=~ f^{a_1a_2b}f^{ba_3a_P}\,B(\D_1-1,\D_2-1,\D_3-1)\d(\D_1+\D_2+\D_3-2-\D_P)~~.
\end{equation}
For the second term, we have the following integrals
\begin{equation}
    \begin{split}
        &\int\,\frac{d\D_q}{2\pi i}\int\,\frac{d\D_l}{2\pi i}\,\left(\prod_{i=1}^3\int_0^{+\infty}d\omega_i\omega_i^{\D_i-1}\right)\,\int_0^{1}dt\,\tilde{f}\left(t;\frac{\omega_1}{\omega_P},\frac{\omega_2}{\omega_P},\frac{\omega_3}{\omega_P}\right)\\
        &\qquad\qquad\qquad\qquad\qquad\qquad\qquad\qquad\qquad\frac{t^{1-\D_q}(1-t)^{1-\D_l}\omega_P^{2-\D_q-\D_l}}{\omega_1\omega_2\omega_3}\,:\oh^{a_q}_{\D_q,-}\oh^{a_l}_{\D_l,-}:\\
        &~=~ \int\,\frac{d\D_q}{2\pi i}\int\,\frac{d\D_l}{2\pi i}\,\int_0^{+\infty}d\omega_P\,\omega_P^{\D_1+\D_2+\D_3-1-\D_q-\D_l-1}\,\left(\prod_{i=1}^3\int_0^1\frac{d\s_i}{\s_i}\s_i^{\D_i-1}\right)\\
        &\qquad\qquad \d(1-\s_1-\s_2-\s_3)\int_0^{1}dt\,\tilde{f}\left(t;\s_1,\s_2,\s_3\right)\,t^{1-\D_q}(1-t)^{1-\D_l}\,:\oh^{a_q}_{\D_q,-}\oh^{a_l}_{\D_l,-}: ~~.\\
    \end{split}
\end{equation}
The $\omega_P$-integral turns into a conformal weight delta function,
\begin{equation}
    \int_0^{+\infty}d\omega_P\,\omega_P^{\D_1+\D_2+\D_3-1-\D_q-\D_l-1} ~=~ (2\pi i)\,\d(\D_1+\D_2+\D_3-\D_q-\D_l-1)~~.
\end{equation}
Finally, the ${\cal D}$ in (\ref{eq:celestialO:OO:}) reads
\begin{equation}\boxed{~
    \begin{aligned}
        &{\cal D}^{a_1^-a_2^-a_3^-}{}_{a_q^-a_l^-}(\D_1,\D_2,\D_3,\D_q,\D_l)\\
        ~=&~ \frac{1}{(2\pi i)}\Big(f^{a_1a_2a_q}\,\d^{a_3,a_l} + f^{a_1a_3a_l}\,\d^{a_2,a_q} \Big)\,\d(\D_1+\D_2+\D_3-\D_q-\D_l-1) \\
        &\times~  \left(\prod_{i=1}^3\int_0^1\frac{d\s_i}{\s_i}\s_i^{\D_i-1}\right)\,\d(1-\s_1-\s_2-\s_3)\int_0^{1}dt\,\tilde{f}\left(t;\s_1,\s_2,\s_3\right)\,t^{1-\D_q}(1-t)^{1-\D_l} ~~,
    \end{aligned}~}\label{eq:someqs}
\end{equation}
with $\tilde{f}$ satisfying 
\begin{equation}
    \int_0^{1}\,dt\, \tilde{f}\left(t;\s_1,\s_2,\s_3\right)  ~=~ 1  ~~.
\end{equation}
If $\tilde{f}$ were a constant, the integral would reduce to a Beta function similar to the ${\cal C}$ coefficient for single-particle operators. 
Moreover generally, introducing the ``Kernel"
\begin{equation}
    K^{\omega_1 \omega_2 \omega_3}_{\quad \omega_q \omega_l} ~:=~ \delta(\omega_q + \omega_l -\omega_1 -\omega_2-\omega_3)\, \frac{{f}(\omega_q;\omega_1,\omega_2,\omega_3)}{\omega_1\omega_2\omega_3}
\end{equation}
the OPE coefficient can be expressed compactly as follows
\begin{equation}
    \begin{aligned}
        &{\cal D}^{a_1^-a_2^-a_3^-}{}_{a_q^-a_l^-}(\D_1,\D_2,\D_3,2-\D_q,2-\D_l)\\
        &\qquad ~=~ \frac{1}{(2\pi i)^2}\,\Big(f^{a_1a_2a_q}\d^{a_3,a_l} + f^{a_1a_3a_l}\d^{a_2,a_q} \Big)~\int \prod_{i=1,2,3,q,l} d\omega_i\,\omega_i^{\Delta_i-1} K^{\omega_1 \omega_2 \omega_3}_{\quad \omega_q \omega_l}~~.
    \end{aligned}
\end{equation}

\section{Closing Remarks}\label{sec:conclusion}

In this paper we showed how to extract multiparticle OPE coefficients from amplitudes. We close by discussing various applications and generalizations that are worth pursuing.

\paragraph{Ward Identities for Celestial Symmetries}
Our previous encounters with multiparticle primaries have been in the context of soft physics: in reproducing the Ward identities in phase space~\cite{Freidel:2021ytz,Freidel:2023gue,Hu:2022txx,Hu:2023geb}, and in understanding the fate of the Jacobi identity for these currents~\cite{Ball:2022bgg,Ball:2023sdz}. For the asymptotic symmetries, this phase space approach explains the universal algebra of matter sector light-ray operators observed in the conformal collider literature\cite{Hofman:2008ar,Cordova:2018ygx}. More importantly for the celestial holography program, antipodal matching of the $in$ and $out$ charges gives us a set of Ward identities for the $\cal S$-matrix that can be explicitly checked within amplitudes and are equivalent to known and new soft theorems~\cite{Strominger:2017zoo}. Meanwhile the more recently discovered $\wedge L w_{1+\infty}$ symmetry was extracted from collinear rather than soft limits~\cite{Guevara:2021abz,Strominger:2021mtt} of amplitudes. The phase space approach shows us how to go beyond the wedge by including additional multiparticle terms and the analogous Ward identities~\cite{Cresto:2024fhd,Cresto:2024mne} would seemingly involve both soft and collinear limits that include multiparticle terms. The results here set up the ground work needed to check this proposal within explicit amplitudes~\cite{fhp}.

One technical thing to note is that the order of the collinear limit and soft limit can be very subtle! For example, consider the following tree-level diagram
\begin{equation}
    \begin{tikzpicture}[baseline={([yshift=-0.9ex]current bounding box.center)},scale=0.9]
        \draw[thick] (-1,0) node[left]{$s$} -- (1,0) node[above]{$\frac{1}{(P_s+P_1)^2}$}  -- (2,0) node[above]{} -- (3,0) node[right]{$\frac{1}{(P_s+P_1+P_2)^2}$};
        \draw[thick] (0,0) -- (0,-1) node[left]{$1$};
        \draw[thick] (2,0) -- (2,-1) node[right]{$2$};
\end{tikzpicture}~~.
\end{equation}
If we take $1||2$ collinear first, then take soft limit for particle $s$, both the $1/(P_s+P_1)^2$ and $1/{(P_s+P_1+P_2)^2}$ propagators give a $1/\omega_s$ singularity. Namely, the total leading soft behavior would be $1/\omega_s^2$, which is problematic.
Identifying the proper prescription that is compatible with both the soft limit and collinear limit is still an open and interesting question. 

\paragraph{Generic Single-particle Exchanges}

Our ansatz~\eqref{equ:OOtoO}-\eqref{eq:O:OO:OPE} captured the celestial OPE coefficients between the single-particle operators and a particular type of two-particle composite operators.  It is natural to want to generalize to higher particle terms, composite primaries built from descendants, and descendants. As we saw for the two-particle case above, the composite operators involving descendants appear at subleading order in the OPE. Meanwhile the higher particle composite operators can give rise to higher order poles in the (anti-)holomorphic collinear limits. 

While the basis for multiparticle celestial operators will be systematically addressed in~\cite{Kulp:2024scx}, one would still want to explicitly compute OPE coefficients within amplitudes or the equivalent boundary correlation functions. Following the amplitude route as we did for the two-particle case here, amounts to look at higher multiplicity collinear limits. As in our discussion above, one expects the higher-point factorization channels to capture the single-particle exchange terms. Namely, all the single-particle terms appearing on the right-hand side of OPEs between multiparticle operators.

\paragraph{Generic Multiparticle Exchanges}
To get the full celestial OPE we of course need to be able to extract terms with multiparticle operators on the right-hand side. In section~\ref{sec:2ptexchange}, we saw how we could constrain 
these coefficients by explicitly evaluating amplitudes in the corresponding collinear limits and collecting terms. To get the full celestial OPE we of course need to be able to extract terms with multiparticle operators on the right-hand side. In section~\ref{sec:2ptexchange}, we saw how we could constrain these coefficients by explicitly evaluating amplitudes in the corresponding collinear limits and collecting terms. This has certain limitations in fully determining the OPE from 4d scattering and we think more 2d constraints will be needed in future investigations, such as crossing symmetry or positivity. We do not think this is inconsistent with the 4d picture which is completely read off from the amplitude, and does not depend on functions such as $f$ in \eqref{eq:someqs}.

A more systematic way would be to understand the appropriate inner product that we want to use for the celestial multiparticle states in order to lower all the indices on celestial OPE coefficients so that they are on equal footing. A step in this direction for the two-particle states is contained in appendix~\ref{appen:2out}. More generally, one would like to understand the analog of CFT completeness relations from the perspective of 4d unitarity. Computing loop-level corrections to the OPE coefficients computed here would be a relevant next step.

\section*{Acknowledgements}

It is a pleasure to thank Adam Ball,  Laurent Freidel, and Justin Kulp for valuable discussions. AG is supported by the Black Hole Initiative and the Society of Fellows at Harvard University, as well as the Department of Energy under grant DE-SC0007870. 
The research of YH and SP is supported by the Celestial Holography Initiative at the Perimeter Institute for Theoretical Physics and the Simons Collaboration on Celestial Holography. Research at the Perimeter Institute is supported by the Government of Canada through the Department of Innovation, Science and Industry Canada and by the Province of Ontario through the Ministry of Colleges and Universities.

\appendix

\section{Multicollinear Limits and Kinematics}\label{appen:conventions}

In this appendix, we summarize the (anti-)holomorphic multicollinear limits and kinematics applied in section~\ref{sec:universal}. Multicollinear limits mostly follow from~\cite{Ball:2023sdz} up to some overall factors from our spinor conventions. 
Kinematics will be derived following the CSW rules~\cite{Cachazo:2004kj}.

First, the spinor helicity variables are parametrized as follows
\begin{equation}
    |\lambda\rangle ~=~ \sqrt{\omega}\begin{pmatrix}
        1\\
        z
    \end{pmatrix} ~~,~~ 
    |\tilde{\lambda}] ~=~ \sqrt{\omega}\begin{pmatrix}
        1\\
        \bz
    \end{pmatrix} ~~.
\end{equation}
Below we list the holomorphic and antiholomorphic 3-collinear limits and derive the kinematics following $\overline{\rm MHV}$ and MHV CSW rules respectively. 

\paragraph{Holomorphic multicollinear limits}

The holomorphic multicollinear limit means that we consider collinear limits between multiple $|\lambda\rangle$ spinors or equivalently $z$ variables while keeping $|\lambda]$ or $\bz$ generic. A generic holomorphic collinear limit between $z_1$, $z_2$, and $z_3$ can be parametrized as follows
\begin{equation}
   \begin{split}
        |1\rangle ~=&~ \sqrt{\omega_1}\,|\hat{p}\rangle + \epsilon\,\eta_1\,\sqrt{\omega_1}\,|r\rangle ~~\leftrightarrow~~ z_1~=~z_p + \epsilon\,\eta_1~~,~~ \\
        |2\rangle ~=&~ \sqrt{\omega_2}\,|\hat{p}\rangle + \epsilon\,\eta_2\,\sqrt{\omega_2}\,|r\rangle ~~\leftrightarrow~~ z_2~=~z_p + \epsilon\,\eta_2 ~~,~~ \\
        |3\rangle ~=&~ \sqrt{\omega_3}\,|\hat{p}\rangle ~~\leftrightarrow~~ z_p~=~z_3 ~~, 
   \end{split}
   \label{equ:123-eta}
\end{equation}
where we choose $|r\rangle=\begin{pmatrix}0\\1\end{pmatrix}$ and take $\epsilon\to 0$ with other parameters fixed. Here we also introduce $\eta$ variables to denote the relative collinearity.

\paragraph{Kinematics for $\overline{\rm MHV}$ vertices} 
As mentioned in the main text, while computing the $s$, $t$, and $u$ contributions (\ref{equ:diagrams}) to the splitting function, we follow the CSW rules~\cite{Cachazo:2004kj}. 
For $\overline{\rm MHV}$ vertices, the kinematics of the internal off-shell gluon $q$ can be derived as follows. First we define the $\tilde{\lambda}_{q,\dot\a}$ by
\begin{equation}
   \tilde{\lambda}_{q,\dot\a} ~:=~ (P_q)_{\a\dot\a}\,\eta^{\a} ~~,~~
   \label{equ:defq]}
\end{equation}
where $P_q$ is the momentum of the off-shell gluon $q$. $\eta^{\a}$ is an arbitrary spinor and we will take it as $|\eta\rangle=\begin{pmatrix}
    0\\
    -1
\end{pmatrix}$ in all diagrams, so that $\lambda_{i,\a}\eta^{\a}=\sqrt{\omega_i}$. Then for each diagram in \eqref{equ:diagrams}, $|q]$ can be written in terms of $|1]$, $|2]$, and $|3]$ by using the momentum conservation. For example, in the first diagram, we have
\begin{equation}
    q ~=~ P_1+P_2 ~~.  
\end{equation}
Plugging into (\ref{equ:defq]}) yields 
\begin{equation}
    |q] ~=~ \sqrt{\omega_1}\,|1] +  \sqrt{\omega_2}\,|2] ~~.
\end{equation}
Similarly, for other diagrams we have
\begin{align}
    & q ~=~ P_2+P_3 ~~\Rightarrow~~ |q] ~=~ \sqrt{\omega_2}\,|2] +  \sqrt{\omega_3}\,|3] ~~,~~ \\
    & q ~=~ P_1+P_3 ~~\Rightarrow~~ |q] ~=~ \sqrt{\omega_1}\,|1] +  \sqrt{\omega_3}\,|3] ~~.
\end{align}
The gluon $P$ is also off-shell and thus we define 
\begin{equation}
    \tilde{\lambda}_{P,\dot\a} ~:=~ P_{\a\dot\a}\,\xi^{\a} ~~,
\end{equation}
where $\xi^{\a}$ is an arbitrary spinor and we take it to be the one satisfying $\lambda_{i,\a}\xi^{\a}=\sqrt{\omega_i}/\sqrt{\omega_P}$ with $\omega_P=\omega_1+\omega_2+\omega_3$. Momentum conservation yields  
\begin{equation}
   P~=~P_1+P_2+P_3 ~~\Rightarrow~~ |P] ~=~ \sqrt{\frac{\omega_1}{\omega_P}}\,|1] + \sqrt{\frac{\omega_2}{\omega_P}}\,|2]  +\sqrt{\frac{\omega_3}{\omega_P}}\,|3] ~~.~~ 
    \label{eq:|P]}
\end{equation}

\paragraph{Anti-holomorphic multicollinear limits}

For anti-holomorphic multicollinear limits, we have an analogous setup to the one above, where we just exchange $z \leftrightarrow \bz$ and $|~\rangle \leftrightarrow |~]$. Namely, a generic anti-holomorphic 3-collinear limit can be parametrized as follows
\begin{equation}
   \begin{split}
        |1] ~=&~ \sqrt{\omega_1}\,|\hat{p}] + \bar{\epsilon}\,\bar{\eta}_1\,\sqrt{\omega_1}\,|r] ~~\leftrightarrow~~ \bz_1~=~\bz_p + \bar{\epsilon}\,\bar{\eta}_1~~,~~ \\
        |2] ~=&~ \sqrt{\omega_2}\,|\hat{p}] + \bar{\epsilon}\,\bar{\eta}_2\,\sqrt{\omega_2}\,|r] ~~\leftrightarrow~~ \bz_2~=~\bz_p + \bar{\epsilon}\,\bar{\eta}_2 ~~,~~ \\
        |3] ~=&~ \sqrt{\omega_3}\,|\hat{p}] ~~\leftrightarrow~~ \bz_p ~=~ \bz_3 ~~. 
   \end{split}
   \label{equ:bz123-eta}
\end{equation}

\paragraph{Kinematics for MHV vertices}
The kinematics for MHV vertices can be derived similarly as $\overline{\rm MHV}$ vertices. Now for the off-shell gluons $q$ and $P$ we define
\begin{equation}
   \lambda_{q,\a} ~:=~ (P_q)_{\a\dot\a}\,\tilde{\eta}^{\dot\a} ~~,~~
   \lambda_{P,\a} ~:=~ P_{\a\dot\a}\,\tilde{\xi}^{\dot\a} 
   \label{equ:defqrangle}
\end{equation}
and choose the spinors $\tilde{\eta}^{\dot\a}$ and $\tilde{\xi}^{\dot\a}$ to be the ones yielding $\tilde{\lambda}_{i,\dot\a}\tilde{\eta}^{\dot\a}=\sqrt{\omega_i}$ and $\tilde{\lambda}_{i,\dot\a}\tilde{\xi}^{\dot\a}=\sqrt{\omega_i}/\sqrt{\omega_P}$ in all diagrams. Momentum conservation then gives the following expressions 
\begin{align}
    & q ~=~ P_1+P_2 ~~\Rightarrow~~ |q\rangle ~=~ \sqrt{\omega_1}\,|1\rangle +  \sqrt{\omega_2}\,|2\rangle ~~,~~ \\
    & q ~=~ P_2+P_3 ~~\Rightarrow~~ |q\rangle ~=~ \sqrt{\omega_2}\,|2\rangle +  \sqrt{\omega_3}\,|3\rangle ~~,~~ \\
    & q ~=~ P_1+P_3 ~~\Rightarrow~~ |q\rangle ~=~ \sqrt{\omega_1}\,|1\rangle +  \sqrt{\omega_3}\,|3\rangle ~~,
\end{align}
and 
\begin{equation}
   P~=~P_1+P_2+P_3 ~~\Rightarrow~~
   |P\rangle ~=~ \sqrt{\frac{\omega_1}{\omega_P}}\,|1\rangle + \sqrt{\frac{\omega_2}{\omega_P}}\,|2\rangle  +\sqrt{\frac{\omega_3}{\omega_P}}\,|3\rangle ~~,~~ 
    \label{eq:|Prangle}
\end{equation}
where $\omega_P=\omega_1+\omega_2+\omega_3$.  

\section{Multiparticle Inner Product}\label{appen:2out}
Here we have extracted celestial OPE coefficients between the single-particle and two-particle states. As noted above, until we understand the inner product the operators appearing on the left- and right-hand sides of the OPE are on different footing. This is particularly subtle for celestial CFT where, in contrast to ordinary Euclidean CFTs, the low point functions are distributional~\cite{Pasterski:2017ylz}. This can be mitigated with appropriate shadow- or light-transforms~\cite{ss,Fan:2021isc,Sharma:2021gcz,Hu:2022syq,De:2022gjn,Jorge-Diaz:2022dmy} and in~\cite{Crawley:2021ivb,Cotler:2023qwh} an inner product reproducing both the analytic form and the notion of adjoint used in radial quantization was worked out for the single-particle states. Here we will comment on how to generalize this to the multiparticle case. For simplicity, we focus on the case scalar fields in this appendix. 

The single-particle inner product that reproduces the correct radial quantization conjugation is the RSW inner product, which involves a reflection in $X^3$ mapping the north to south poles, a shadow transform on the $out$ operators and a Weyl reflection on the weights
\be
\llangle \Delta' |\Delta\rangle~=~\lim_{z\rightarrow\infty} |z|^{2\Delta'} \langle0|\widetilde O_{\Delta'}(z,\bz)O_{\Delta}(0,0)|0\rangle~=~\frac{\Gamma(2-\Delta)}{\Gamma(\Delta-1)}2^5 \pi^3{\boldsymbol{\delta}}(i(\Delta'-\Delta)) ~~.
\ee
Its relation to 4D inversions is explored in~\cite{Jorstad:2023ajr}. It has the effect of mapping the momentum space single-particle inner product 
\be
\langle p| p' \rangle ~=~ (2\pi)^3 \,2\omega\, \delta^{(3)}(p-p') 
\ee
to one that matches what we expect for the radial quantized theory. 
Mechanically this works out due to the fact that the two-point function of our Mellin transformed states is distributional, while the two-point function has a power law form
\be
\langle\widetilde O_{\Delta_1}(z_1,\bz_1) O_{\Delta_2}(z_2,\bz_2)\rangle~=~\frac{\Gamma(2-\Delta_1)}{\pi\Gamma(\Delta_1-1)}\frac{2(2\pi)^4}{|z_1-z_2|^{2\Delta_1}}{\boldsymbol{\delta}}(i(\Delta_1-\Delta_2))
\ee
where, for clarity, we should mention that $\Delta_1$ is the weight after taking a shadow of a weight $2-\Delta_1$ operator and the distribution $\boldsymbol{\delta}$ is defined for weights off the principal series in~\cite{Donnay:2020guq}. This is the opposite of what is expected in usual 2D CFTs, and one needs to be careful in applying the shadow formalism of~\cite{Ferrara:1972ay,Ferrara:1972uq,Ferrara:1972xe,Ferrara:1972kab} 
where the projectors \cite{SimmonsDuffin:2012uy}
\begin{equation}
\int d^dx \,\mathcal{O}(x)|0\rangle\langle0|\tilde{\mathcal{O}}(x)
\end{equation}
are used to extract exchanges involving the conformal multiplets of the primary operator $\mathcal{O}$ and its shadow.

For our purposes we are just worried about defining the radially quantized $out$ states. The $n$-particle states have even more delta functions
\be
\langle p_1\cdots p_n| p_1'\cdots p_n'\rangle ~=~ \prod_{i=1}^n\,  (2\pi)^3\, 2\omega_i\, \delta^{(3)}(p_i-p_i') ~+~ {\rm perm} ~~, 
\ee
where the last term indicates permutations of the $i'$ when we have identical particles.
As such, to get an analytic two-point function for our multiparticle operators in the free theory we will need to shadow-transform each of the particles. Namely, our two-particle $out$ state is
\be
\llangle :O_{1'}O_{2'}:| ~=~
\lim\limits_{z\rightarrow\infty} |z|^{2\Delta_1'+2\Delta_2'}\, \widetilde{O}_{\Delta_1'}(z,\bz)\,\widetilde{O}_{\Delta_2'}(z,\bz)
\ee
giving the inner product with the two-particle $in$ state $|:O_1O_2:\rangle=~:O_1O_2:|0\rangle$
\be\badat{3}
\llangle :O_{1'}O_{2'}:|:O_{1}O_{2}:\rangle~&=~\frac{\Gamma(2-\Delta_1)}{\Gamma(\Delta_1-1)}\frac{\Gamma(2-\Delta_2)}{\Gamma(\Delta_2-1)}\,2^{10}\pi^6\\
 &\times\Big[{\boldsymbol{\delta}}(i(\Delta_{1'}-\Delta_1)){\boldsymbol{\delta}}(i(\Delta_{2'}-\Delta_2))-{\boldsymbol{\delta}}(i(\Delta_{2'}-\Delta_1)){\boldsymbol{\delta}}(i(\Delta_{1'}-\Delta_2))\Big]
\eadat\ee
which is non-distributional on the celestial sphere.

\bibliographystyle{utphys}
\bibliography{reference}

\end{document}